\documentclass[aps,twocolumn,amssymb]{revtex4}

\usepackage{graphicx}

\usepackage{color}

\def\bea{\begin{eqnarray}}
\def\eea{\end{eqnarray}}
\def\ben{\begin{equation}}
\def\een{\end{equation}}
\def\benu{\begin{enumerate}}
\def\enu{\end{enumerate}}

\def\n{n}

\def\sss{\scriptscriptstyle\rm}

\def\1var{(\bx_1...\bx\N)}

\def\br{{\bf r}}

\def\bx{{\br t}}

\def\x{_{\sss X}}
\def\c{_{\sss C}}
\def\s{_{\sss S}}
\def\xc{_{\sss XC}}
\def\Hx{_{\sss HX}}

\def\Hxc{_{\sss HXC}}

\def\N{_{\sss N}}
\def\H{_{\sss H}}

\def\unif{^{\rm unif}}

\def\ALDA{^{\rm ALDA}}

\def\sph_int{ {\int d^3 r}}

\def\sig{_{\sigma}}
\def\sigp{_{\sigma'}}

\begin{document}
\renewcommand{\thefootnote}{\fnsymbol{footnote}} 
\renewcommand{\theequation}{\arabic{section}.\arabic{equation}}

\title{Time-dependent density functional theory of high excitations:  To infinity, and beyond}

\author{Meta van Faassen$^a$}
\email[]{E-mail: m.van.faassen@few	.vu.nl} 
\author{Kieron Burke$^b$}

\affiliation{$^a$ Afdeling Theoretische Chemie, Scheikundig Laboratorium der Vrije Universiteit, De Boelelaan 1083, NL-1081 HV Amsterdam, The Netherlands\\
$^b$Departments of Chemistry and of Physics, University of California, Irvine, CA 92697, USA}

\date{\today}

\begin{abstract}
We review the theoretical background for obtaining both quantum defects and scattering phase shifts from time-dependent density functional theory. The quantum defect on the negative energy side of the spectrum and the phase shift on the positive energy side merge continuously at $E=0$, allowing both to be found by the same method. 
We illustrate with simple one-dimensional examples: the spherical well and the delta well potential. As an example of a real system, 
we study in detail elastic electron scattering from the He${}^{+}$ ion. We show how the results are influenced by different approximations to the unknown components in (time-dependent) density functional theory: the ground state exchange-correlation potential and time-dependent kernel. We also revisit our previously obtained results for $e$-H scattering.
Our results are remarkably accurate in many cases, but fail qualitatively in others.
\end{abstract}

\maketitle

\tableofcontents

\section{Introduction}
\label{intro}
Density functional theory (DFT) \cite{HK64,KS65,DG90} has become popular for calculating ground-state properties of large molecules; it replaces the interacting many-electron problem with an effective single-particle problem that is computationally inexpensive
 to solve. Based on rigorous theorems\cite{HK64,KS65,L82}, and a hierarchy of increasingly accurate approximations, DFT calculations are common in almost all branches of chemistry.

In principle, all properties, including excitations, are
functionals of the ground-state density, but accurate, general-purpose
approximations to obtain these excitations have yet to be found.  Instead, 
time-dependent density functional theory (TDDFT), based on 
another theorem~\cite{RG84}, provides a usefully accurate
approach for many applications~\cite{EBF07}.  In particular,
low-lying single excitations are often well-approximated, and
of great interest in many applications.  Moreover, the 
computational technology for solving the necessary equations already existed,
so TDDFT was rapidly built into existing codes and has become popular for calculating excitations.
But TDDFT is mostly used to extract bound-bound transitions,
and even then, only low-lying single excitations.  In this
perspective, we review our work extending TDDFT to higher
excitations, including both Rydberg states and the continuum,
i.e., $n\to\infty$ and beyond.

We begin with a review of the theory of the quantum defect and scattering phase shift aimed primarily at our electronic-structure readers. Next, we give some background on (TD)DFT and how to obtain the quantum defect and scattering phase shift from a (TD)DFT calculation. We prove that within TDDFT it is allowed to use the method of obtaining pseudo continuum states from a hard wall cavity, and thus obtain the phase shifts from a ``bound'' state calculation. We show results for the quantum defect of He and the $e$-He${}^{+}$ scattering phase shift.

\section{Theoretical background}
\label{theo}
The theoretical ingredients needed to understand this work originate from
two distinct fields, scattering theory and density functional theory.
While closely related in principle, these areas are almost mutually
exclusive in practice.  A reader from either field might skip the
familiar material.

\subsection{Scattering theory and Rydberg states}
\label{background}
The electronic structure of atoms is understood in a rather simple way if one of the electrons is in a highly excited state. In that case we consider the system as consisting of the excited electron in a local potential, originating from the nucleus and all other (tightly) bound electrons. This local potential is a long-ranged Coulomb potential for neutral and positively charged atoms  and  a shorter ranged potential  for negatively charged atoms.

An atom with a highly excited electron in a long-ranged atomic potential is also known as a {\em Rydberg} atom. Such atoms are highly studied and an overview of their properties can for example be found in Refs.~\cite{G94,CPW05}, and references therein. Since Rydberg atoms can be regarded as one electron in an effective local potential, they are similar to the hydrogen atom. The difference between Rydberg atoms and hydrogen atoms is expressed in terms of the quantum defect, which we introduce in the next section.

The infinite Rydberg series merges into the continuum at the ionization limit of the system; it is this boundary between the finite states and the continuum that we explore in detail in this paper. If the electron resides in a continuum state instead of a Rydberg state, it scatters from
the resulting ionic core. We explain this further in section~\ref{phase}.

\subsubsection{The quantum defect and its smoothness}
\label{QD}
 For any spherical one-electron potential that decays as $-1/r$ at
large distances, the bound-state transitions form a Rydberg series.
The high-energy levels of such a system are given
by the Rydberg equation, which in atomic units reads,
\ben
E^{i}_{n}=I-\frac{1}{2(n-\mu_i)^2},
\label{e:qd}
\een
where $i=(nl)$ and $E^{i}_{n}$ is
the excitation energy to a state of principal quantum number $n$ and
angular momentum $l$, and $I$ is the ionization energy of the system studied. We use atomic units throughout.
The quantum defect, $\mu_{i}$, is purely determined by that part of the potential that
differs from $-1/r$.

As is shown in Eq.~(\ref{e:qd}), the quantum defect behaves differently for different angular momentum
states.  For high angular momentum the quantum defect is generally
small, since the shielding effect of the core is mostly maintained. 
The existence of a quantum defect requires the atomic potential to decay like $-1/r$,
producing a Rydberg series, but its value is determined by the interior region of the potential~\cite{F81}.

\begin{figure}[tbp]
\includegraphics[width=7cm]{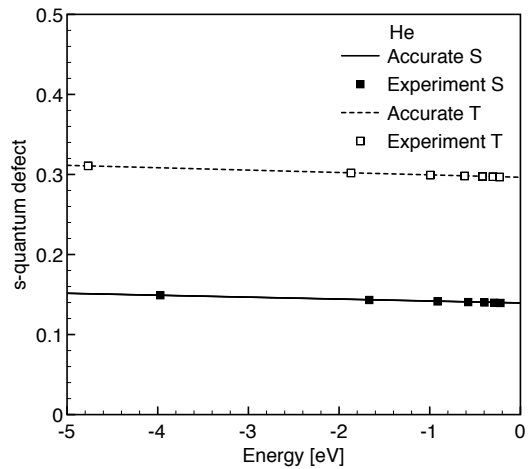}
\caption{\label{fig:HeNISTDrake}Accurate quantum defects for the He atom from Ref.~\cite{D96} (Exact)
and experimental data from NIST~\cite{NIST}.}
\end{figure}
In Fig.~\ref{fig:HeNISTDrake} we show the exact quantum defects from high level wave function calculations~\cite{D96} , fitted to a smooth function as explained below, and we compare them with the experimental values~\cite{NIST}. We notice that both in the singlet and in the triplet case the quantum defect is a smooth function of energy,
so we
 fit it to the following expansion,
\ben\label{eq:fit}
\mu^{(p)}(E) =\sum_{i=0}^p \mu_i E^i,\quad E=\omega-I.
\een
In Fig.~\ref{fig:HeNISTDrake}, the fits are given by $\mu^{(1)}(E) = 0.1395 - 0.0655 E$ for singlet and $\mu^{(1)}(E) = 0.2965 - 0.0811 E$ for triplet.
In previous work~\cite{FB06,FBb06,F06} we showed that in (TD)DFT this
fit works well and we only need two or three coefficients to describe
all the data accurately. The reason for fitting the quantum defect is that it is hard to calculate the higher Rydberg states. If we have a way to fit the lower quantum defects accurately, we instantly have a way to obtain the values of {\em all} higher lying Rydberg states, up to the continuum. The coefficients for the helium atom can be found in the references above and are not repeated in this paper.

\subsubsection{Potential scattering and the phase shift}
\label{phase}
\begin{figure}[tbp]
\includegraphics[width=7cm]{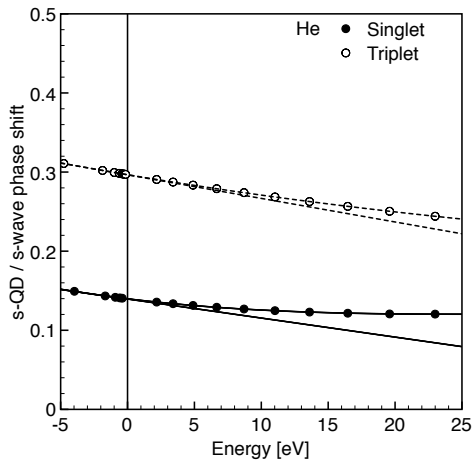}
\caption{\label{fig:HeQDposneg}Helium atom negative energy singlet and triplet s-quantum defects from Ref.~\cite{D96}
and positive energy results from Ref.~\cite{B02}. The phase shifts are divided by $\pi$ so they match with the quantum defects at $E=0$.
The coefficients corresponding to the fits are reported in the text.}
\end{figure}
The quantum defect, discussed in the previous section, is not only an important quantity for studying bound Rydberg states, but
it is also relevant for scattering states. Seaton's theorem~\cite{S58} tells us
that the quantum defect is a smooth function of energy as $E\to 0$,
and it merges continuously with the phase-shift (relative to pure Coulomb
scattering),
\begin{equation}\label{e:seaton}
\delta_{l}(0)=\pi\mu_{l}(0). 
\end{equation}
We illustrate this in Fig.~\ref{fig:HeQDposneg}
for the case of the helium atom. 
In this figure, and all that follow, we plot $\pi\mu_{l}$ instead of the quantum defect itself. In this graph we also show a continuous fit of the points. Since the positive energy part is more curved than the QDs alone, we need more coefficients for an accurate fit. The fits are given by,  $0.1399-0.0528E+ 0.0358 E^{2}$ for the singlet case and   $0.2967-0.0764 E+ 0.0167 E^{2}$  for the triplet case. Together with these fits we also show the straight line fit of Fig.~\ref{fig:HeNISTDrake}. It is clear that this fit is good for negative and very small energies, but diverges for larger energies.
Eq.~(\ref{e:seaton}) is only relevant if there is a quantum defect to consider, i.e. if we consider an atom with a long-ranged potential. Such a long-ranged atomic potential can in general be decomposed into a pure Coulomb ($-1/r$) and a short-ranged part,
\begin{equation}\label{eq:coulsr}
v_{\rm sr}(r)=-\frac{1}{r}-v(r).
\end{equation}
But to understand its scattering properties, we must first review elementary scattering from a spherical potential.

If a potential is short-ranged, the differential scattering cross section is given by~\cite{F98}
\begin{equation}
\frac{d\sigma}{d\Omega}=|f(\theta)|^{2}.
\end{equation}
The scattering amplitude is
written in terms of partial waves,
\begin{equation}\label{e:fexpand}
f(\theta)=\sum_{l=0}^{\infty}f_{l}P_{l}(\cos\theta),
\end{equation}
where $P_{l}(\cos\theta)$ is a Legendre polynomial
and the partial amplitude can be expressed in terms of phase shifts:
\begin{equation}
f_{l}=\frac{2l+1}{k}e^{i\delta_{l}}\sin\delta_{l}.
\end{equation}
It follows from this equation that if we have the phase shift for all values of $l$ we can calculate the differential cross section. In practice the few lowest $l$ values will often be dominant and almost completely determine the cross section.

For a long-ranged potential the equations are more complicated. In the  case of atoms,
we can separate the total potential into a Coulomb plus a short range part. The differential cross section is then given by~\cite{F98}
\begin{equation}\label{e:scattamp}
\frac{d\sigma}{d\Omega}=|f_{C}(\theta)+f_{\rm sr}(\theta,\phi)|^{2},
\end{equation}
where $f_{C}(\theta)$ is the pure Coulomb amplitude and $f_{\rm sr}(\theta,\phi)$ is the additional scattering amplitude as a result of the short-range part of the potential.
This is necessary because $f_{C}(\theta)$ diverges as $\theta\to 0$, making the cross-section itself divergent. Thus, scattering from a long-ranged potential is qualitatively different from that of a short-ranged potential.
The additional scattering amplitude of Eq.~(\ref{e:scattamp}) does not depend on the angle $\phi$ and can again be expanded as Eq.~(\ref{e:fexpand}). The partial wave amplitudes are given by,
\begin{equation}
f_{{\rm sr},l}=\frac{2l+1}{k}e^{i\sigma_{l}}e^{i\delta_{l}}\sin\delta_{l}
\end{equation}
where the $\sigma_{l}$ denote the Coulomb phase shifts and the $\delta_{l}$ the phase shifts due to the short-ranged part of the potential.
If the short-range potential is small compared to the Coulomb potential,  the additional amplitude can be calculated using, for example, a distorted wave Born approximation~\cite{F98}. 
In the next section, we explain how we obtain accurate phase shifts in practice.

\subsubsection{Continuum states from spherical boxes}
A useful method we recently (re)discovered~\cite{F35},
and which circumvents the problem of working with
 continuum states, is to put the system of scattering electron and target in a spherical box. In that way a finite number of ``continuum'' states is obtained exactly, namely those that have nodes at the radius where we placed the wall, $r=R_{b}$. 
Assuming one has the liberty to choose any wall radius, {\em all} continuum states can be obtained in this way.  Only for very low-energy continuum states does one need large boxes.

Just as in the previous section, we distinguish two kinds of potentials, short-ranged potentials that effectively vanish after some finite radius $R_{c}$,
and long-ranged potentials that do not. The first kind we studied in Ref.~\cite{FWEZ07} where we investigated the scattering of an electron by the hydrogen atom. The $(N_{T}+1)$-electron system, H${}^{-}$, is a negative ion with a short-ranged potential. Its phase shift is obtained from knowledge of the spherical Bessel functions as a function of the energies and the wall radius, explicitly,
\begin{equation}\label{e:srphase}
\tan\delta_{nl}=-\frac{j_l(k_nR_b)}{n_l(k_nR_b)},
\end{equation}
where $j_{l}$ and $n_{l}$ are the spherical Bessel and Neumann functions.

\begin{figure}[tbp]
\includegraphics[width=7cm]{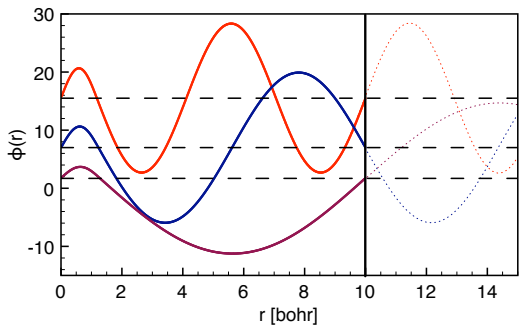}
\caption{\label{f:spherorbs}The lowest three positive energy orbitals for a spherical well within a box with $R_{b}=10$ (radius 1.2, well-depth 3). The dashed curves indicate how the continuum functions continue. The curves are offset for clarity, each zero being marked by a dashed horizontal line. Color figure online.}
\end{figure}
As an example, we show the lowest three of such states for a spherical well in Fig.~\ref{f:spherorbs}, where the wall is placed at $R_{b}=10$. The spherical well potential is given by
\begin{equation}
V(r)= - D\Theta(a-r),
\end{equation}
where $D$ is the depth of the well, $a$ the radius, and $\Theta$ the Heaviside step function.
 The radius $a$ of the spherical well is 1.2 bohr and the depth $D=3$ Hartree. It is clear from this graph that the orbitals are the continuum functions, except that they do not continue beyond $R_{b}=10$ and that there is only a finite number of them.

In this paper we also study electron scattering from He${}^{+}$. The $(N_{T}+1)$-electron system is the neutral helium atom, which has a long-ranged modified Coulomb potential. We use the fact that this potential can be separated in a pure Coulomb potential ($-1/r$) and a short-ranged potential, i.e. that vanished at some finite $r$. All we need to know to obtain the scattering phase shifts are the Coulomb functions as a function of energy and wall radius, so Eq.~(\ref{e:srphase}) becomes
\begin{equation}\label{e:lrphase}
\tan\delta_{nl}=-\frac{F_l(\eta,k_{n}R_b)}{G_l(\eta,k_{n}R_b)},
\end{equation}
where $F_{l}$ and $G_{l}$ are the regular and irregular Coulomb functions. The only requirement on the location of the wall is that it is outside the short-ranged part of the potential. We can place the wall relatively close to the atom, even though we are dealing with a long-ranged potential.

\begin{figure}[tbp]
\includegraphics[width=7cm]{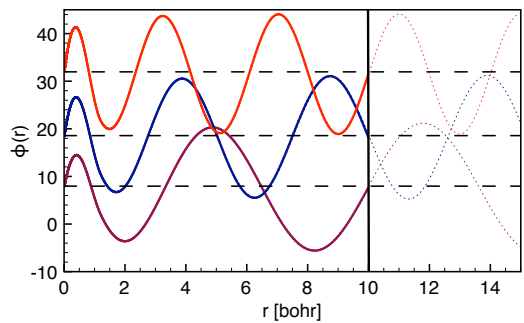}
\caption{\label{f:sphercoulorbs}Same as Fig.~\ref{f:spherorbs}, but with $-1/r$ added to the potential. Color figure online.}
\end{figure}
As an example, we show the lowest three of such states for the same spherical well as above, plus a Coulomb potential, in Fig.~\ref{f:spherorbs}. The total potential is given by,
\begin{equation}
V(r)=-D\Theta(a-r)-\frac{1}{r}.
\end{equation}
Again the wall is placed at $R_{b}=10$. It is again clear that the functions behave exactly like continuum functions.

\subsection{Ground-state density functional theory (DFT)}
We now review briefly both ground-state and time-dependent density
functional theory. We emphasize throughout the distinction between exact DFT and the approximations invariably used in practice.

Density functional theory is based on the theorem by Hohenberg and Kohn (HK)~\cite{HK64}.
This theorem states that, for
a given non-degenerate ground-state density
of Fermions, $\n(\br)$, with a given interaction,
the external potential that produces the density is unique (up to an additive constant).
Thus, if the density is known, the external potential is also known, completely defining the Hamiltonian, i.e., we can obtain all properties of the system from the density alone.
The HK theorem also holds for degenerate ground-states~\cite{L82}
and there exists an analogous theorem that can be applied to spin densities~\cite{BH72,PR72}.

The power of DFT is in the idea of Kohn and Sham (KS) to apply the HK theorem to a system of non-interacting particles. The HK theorem states that the KS potential, which reproduces
the true density of the system, is unique. KS map the fully interacting problem to a non-interacting one with the same density. The KS equations are a set of one-particle Schr{\"o}dinger equations,
\begin{equation}\label{eq:KS}
\left(-\frac{1}{2}\nabla^{2}+v\s(\br)\right)\phi_{i}(\br)=\varepsilon_{i}\phi_{i}(\br),
\end{equation}
where the $\phi_{i}$ are the KS orbitals and $\varepsilon_{i}$ the corresponding orbital energies. The effective KS potential is given by
\begin{equation}\label{eq:vs}
v\s(\br) = v_{\rm ext}(\br) + \int d\br'\frac{n(\br')}{|\br-\br'|} + v\xc(\br).
\end{equation}
The unknown part in this equation, $v\xc(\br)$, is the exchange-correlation (xc) potential, that needs to be approximated in practice. When the exact xc-potential is known, the exact interacting density of the system can be obtained from the KS orbitals,
\begin{equation}
n(\br)=\sum_{i=1}^{N}|\phi_{i}(\br)|^{2},
\end{equation}
where the sum runs over all occupied orbitals.

\begin{figure}[tbp]
\includegraphics[width=7.7cm]{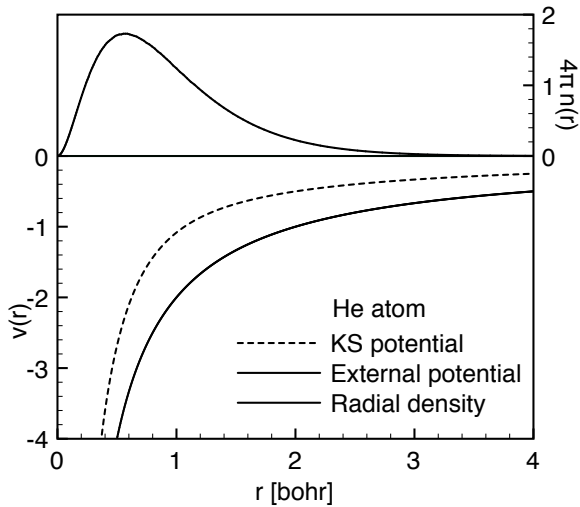}
\caption{\label{f:He_potcompare}Top panel: extremely accurate radial density for the helium atom found via the QMC method~\cite{UG94}.  Bottom panel: The external and KS potentials for the helium atom.}
\end{figure}
We illustrate the KS idea in Fig.~\ref{f:He_potcompare} where we show both the exact KS potential of the helium atom~\cite{UG94} and the external potential from an accurate wave function calculation. Umrigar and Gonze~\cite{UG94} obtained the exact KS potential from an extremely accurate quantum Monte-Carlo calculation for the ground-state of He, calculating the density, and finding the KS potential by inverting the KS equation for the doubly-occupied $1s$ orbital:
\begin{equation}
v\s(r)=\varepsilon_{1s}-\nabla^{2}\phi_{1s}(r)/{\phi_{1s}(r)},
\end{equation}
where $\phi_{1s}(r)=\sqrt{2n(r)}$.
The KS potential is quite different from the external potential. This is because two non-interacting ``electrons'' in the KS potential reproduce the exact electron density of the interacting system, by definition.  For example, because electron-electron repulsion pushes density outward, the KS potential is shallower than the external potential.

The energy of the highest occupied molecular orbital (HOMO) of the exact He KS-potential is $-24.592$eV. This is exactly minus the ionization energy of the helium atom; an illustration of the exactness of Koopman's theorem in DFT~\cite{PPLB82}. This exact relation is a remarkable fact, since the KS system is a fictitious system, but it follows from the exactness of $n(r)$ as $r\to\infty$. In fact, $v\xc(r)\to -1/r$ as $r\to\infty$, for all spherical systems whose density $n(r)$ decays exponentially. Thus, since,
\begin{equation}
v\H(r)=\int dr' \frac{n(\br')}{|\br-\br'|}\to \frac{N}{r} \quad r\to\infty,
\end{equation}
then $v\s(r)\to-1/r$ if $N=Z$, as in Fig.~\ref{f:He_potcompare}, or more rapidly if $N=Z+1$.

In practice, $v\s(\br)$ is not known exactly. DFT calculations use approximations to $E\xc[n]$, whose derivative yields an approximation to $v\xc(\br)$. Many approximate xc-energy functionals have been developed since the start of DFT. The simplest of these, the local density approximation (LDA), gives remarkably good results~\cite{KS65}. It is improved by including the gradient of the density, leading to the generalized gradient approximation (GGA). Some popular GGAs are BLYP (B88~\cite{B88} for exchange and LYP~\cite{LYP88} for correlation) and PBE~\cite{PBE96}. Hybrid potentials are popular among chemists because of their improved accuracy for main-group thermochemistry and transition-state barriers. These hybrid functionals mix in some fraction of exact exchange with a GGA. The most widely used hybrid, B3LYP, contains 3 experimentally fitted parameters~\cite{B93,LYP88,MSSP89}, whereas PBE is derived from general principles of quantum mechanics~\cite{PBE96}.

\begin{figure}[tbp]
\includegraphics[width=7.7cm]{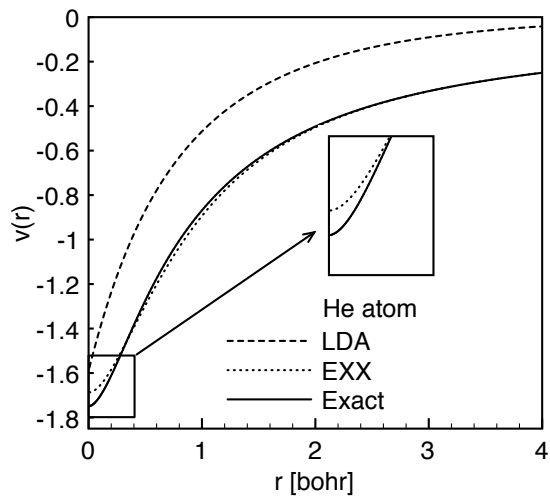}
\caption{\label{f:He_LDAexact}KS potentials for the helium atom: LDA, exchange only (EXX), and exact.}
\end{figure}
A less well-known feature to users of ground-state DFT is that although their favourite approximations often yield very good energies (and therefore structures, vibrations, thermochemistry, etc.) and rather good densities, they have badly behaved potentials, at least far from nuclei. The LDA potential, for example, decays much too fast. We show this in Fig.~\ref{f:He_LDAexact} for the helium atom.
Although the relative energies of low lying states might be accurate, the high lying Rydberg states, in which we have particular interest, are poorly reproduced, if bound at all. In the case of the H$^{-}$ anion, the LDA even fails to bind this system at all. 
These errors are called self-interaction, because $E\x=-E\H$ and $E\c=0$ for one electron, but not with these approximations.
Over the past decade, the technology for treating orbital-dependent functionals, such as the Fock integral (so-called exact exchange), has been developed, and such functionals 
help cure this problem\cite{E03}. This is called 
the optimized effective potential method (OEP)\cite{TS76,KLI92,G96}.
The OEP
method can handle any orbital-dependent functional including simple exchange, and then produces an asymptotically correct $v\s(r)$.

\subsection{Time-dependent density functional theory}
\label{TDDFT}
Next we give a concise overview of the basics of time-dependent density functional theory (TDDFT), focussed on obtaining excitation energies. For a more in-depth review, we refer to Ref.~\cite{EBF07}, and references therein.

\subsubsection{General formalism}
Ground-state DFT does not produce the exact transition energies. The reason is that the KS orbital energies do not correspond to the true bound-state energies. A popular way to obtain the true transition energies is via time-dependent density-functional theory (TDDFT). 
TDDFT is an in principle exact method based on the Runge-Gross theorem~\cite{RG84}. This theorem states that, for time-dependent problems,  the time-dependent potential, and consequently all other properties, is a functional of the time-dependent density.
Armed with this theorem, time-dependent Kohn-Sham (TDKS) equations are defined analogues to Eq.~(\ref{eq:KS}) but with $i\partial\phi_{i}/\partial t$ on the right. These equations describe a system of non-interacting electrons that evolve in a time-dependent Kohn-Sham
potential and produce the same time-dependent density as that of the interacting system of interest, thus replacing the interacting problem with a much simpler non-interacting one, as in the ground-state case. One therefore defines (and needs to approximate) a time-dependent xc-potential, $v\xc[n](\br,t)$, as that defined by Eq.~(\ref{eq:vs}).
 
\subsubsection{Linear response} 
Most practical TDDFT calculations use linear response theory. Within linear response theory, the TDDFT equations simply reduce to an eigenvalue equation~\cite{C95},
\begin{equation}
{\bf{\Omega }}^{(\pm)} {\bf{F}}_i^{(\pm)}  = \omega _i ^2 {\bf{F}}_i^{(\pm)} 
\end{equation}
where the $\omega_i$  are the excitation energies of the system at hand. The superscript $\pm$ means that we have one set of equations for singlet energies, and one for triplet energies.
The key quantity in this equation is obviously the ${\bf{\Omega }}^{\pm}$ matrix, diagonalization of which, when spin-resolved for a
spin-restricted system, immediately yields the transition frequencies $\omega_i$:
\bea\label{e:omegas}
\Omega _{ia,jb}^{(\pm)}  &=& \delta _{ab} \delta _{ij} ( \varepsilon_a - \varepsilon_i)^2  \\\nonumber
&+& 2\sqrt {\varepsilon_a - \varepsilon_i} \left( K_{ia,jb}^{ \uparrow  \uparrow }  
\pm K_{ia,jb}^{ \uparrow  \downarrow }  \right)\sqrt {\varepsilon_b - \varepsilon_j} 
\eea
where the $\varepsilon_i$ are occupied Kohn-Sham (KS) orbital energies and $\varepsilon_a$ are unoccupied KS orbital energies, the ${\bf K}$ matrix elements are discussed below, $\uparrow$ and $\downarrow$ specify the spin orientation. The Kohn-Sham orbital energies correspond to a system of non-interacting electrons, the sum of which, in an exact DFT calculation, gives the true density of the system. The first term of the ${\bf{\Omega }}^{(\pm)}$ is the {\em KS transition energy}. The second term in both matrices corrects the KS result to give the true interacting result, the singlet and triplet transition energies respectively. 

An approximation to solving the above eigenvalue equation is the small matrix approximation (SMA), where we only include diagonal elements of the ${\bf{\Omega }}^{(\pm)}$-matrix~\cite{PGG96,VOC99}.  This leads to a singlet triplet splitting but it does not allow different transitions to mix with each other. The SMA is useful for understanding results, but usually is not applied to obtain the transition energies for practical systems.

\begin{figure}[tbp]
\includegraphics[width=7.7cm]{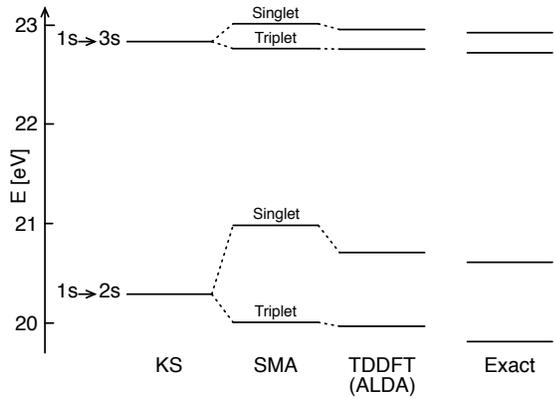}
\caption{\label{f:helevelscompare}The first two transitions for He. Exact KS orbital energy differences, SMA singlet-triplet excitation energies and full TDDFT (ALDA) excitation energies (both evaluated for the exact ground-state KS potential), and the exact values.}
\end{figure}
We illustrate how the transition energies evolve from KS to SMA to full TDDFT in Fig.~\ref{f:helevelscompare}, where we consider the $1s\to 2s$ and $1s\to 3s$ transitions in the helium atom. From a ground-state KS calculation we obtain just the KS orbital energy differences, which are somewhere in-between the exact values. The next step is to use the small matrix approximation (SMA), where we see that especially the singlet values still overestimate the exact values. The final step is to include the full matrix, allowing full mixing of all states.  The full TDDFT results are close to the exact values, but do not match them, because we used an approximate kernel (see below). Of course, all this information, and much more, is beautifully condensed  into the quantum-defect plots throughout this paper. For example, the exact results in Fig.~\ref{f:helevelscompare} are represented by the leftmost points of Fig.~\ref{fig:HeNISTDrake}.

\subsubsection{XC Kernels}
The ${\bf K}$ matrix elements in Eq.~(\ref{e:omegas}) are given by,
\begin{eqnarray}\label{e:kmat}
 K_{ia,jb}^{\sigma \tau } ( \omega ) = 
 \int \!\! \int 
 \frac{\phi _{i\sigma } ( {\bf r} ) \phi _{a\sigma } ( {\bf r} ) \phi _{b\tau } ( {\bf{r'}} )
\phi _{j\tau } ( {\bf{r'}} )}{ [ {\bf r} - {\bf{r'}}] }  \, d{\bf r}d{\bf{r'}} \nonumber \\
+\int \!\! \int \phi _{i\sigma } ( {\bf r} ) \phi _{a\sigma } ( {\bf r} ) f_{\rm{xc}}^{\sigma \tau } ( {\bf r},{\bf{r'}},\omega)\phi _{b\tau } ( {\bf{r'}} )  \phi _{j\tau } ( {\bf{r'}} ) \, d{\bf r}d{\bf{r'}}.
\end{eqnarray}
where we used $\sigma,\tau$ to indicate spins and $i,j$ indicate occupied orbitals and $a,b$ virtual orbitals.
The first term in this equation is the Hartree term, which is only dependent on the ground-state KS orbitals. 
From Eq.~(\ref{e:omegas}) we see that this term cancels in the triplet case (minus combination). The second term contains the {\em xc-kernel}, 
\begin{equation}
f_{{\rm xc}\sigma\sigma' } ( {\bf r},{\bf{r'}},t-t')=\frac{\delta v_{{\rm xc}\sigma}(\br,t)}{\delta n_{\sigma'}(\br',t')}
\end{equation}
whose Fourier transform is simply $f_{\rm{xc}}^{\sigma \tau } ( {\bf r},{\bf{r'}},\omega)$. The xc-kernel is the only part in this expression that needs to be approximated, after the ground-state calculation has been performed.
In this paper we focus on three different kinds of xc-kernels, adiabatic LDA (ALDA), EXX, and hybrid, and study how they perform for the bound and continuum states of the helium atom. All these kernels use the adiabatic approximation, i.e., ignoring the frequency dependence of the xc-kernel.

The simplest xc-kernel is the adiabatic local density approximation (ALDA). The ALDA xc-kernel is given by
\bea
\label{fxcALDA}
&&f_{{\sss XC}\sigma\sigma'}\ALDA[n_0](\br,\br',t,t') =\nonumber\\
&&\delta^{(3)}(\br-\br')\, \delta(t-t')\,
\frac{d^2 e\xc\unif}{dn\sig dn\sigp}
\bigg|_{n\sig = n_{0\sigma}({\bf, r})},
\eea
where $e\xc\unif$ is the xc-density of the uniform electron gas.
The time Fourier-transform of this kernel
has no frequency-dependence
and via a Kramers-Kronig relation, this implies that
it is purely real~\cite{B06}.

Even though finite systems like atoms and molecules are far from representing the homogeneous electron gas, the ALDA is a good approximation in practice. There are systems, though, where the ALDA fails completely. For example the polarizability of long chain molecules is strongly overestimated by the ALDA~\cite{GSGB99}. A solution to this problem is to use orbital dependent functionals~\cite{MWY03,KKP04}, or to solve the time-dependent {\em current} DFT equations with the current dependent Vignale-Kohn functional~\cite{FBLB02}.

In this paper we study the helium atom, in the special case of a two-electron system we can make some other simple choices for the xc-kernel. For two-electrons we can treat exchange exactly, the kernel is simply minus half of the Hartree kernel,
\ben\label{eq:fexx}
f_{{\sss X}\sigma\sigma'}^{\rm EXX}=-\frac{\delta_{\sigma\sigma'}}{|{\bf r}-{\bf r'}|}.
\een
This kernel can be improved by combining with the correlation part of
ALDA. This results in the so-called hybrid xc-kernel,
\ben\label{eq:fhyb}
f_{{\sss XC}\sigma\sigma'}^{\rm hyb}=-\frac{1}{2}\frac{\delta_{\sigma\sigma'}}{|{\bf r}-{\bf r'}|}+\frac{1}{2}f_{{\sss C}{\sigma\sigma'}}^{{\rm ALDA}},
\een
which improves on EXX for the helium atom~\cite{BPG00}.

\section{One-dimensional illustrations}
In this section we treat 
potentials of finite range in one dimension, i.e., $v(x)=0$ beyond some $a$.  We restrict
ourselves to symmetric potentials ($v(-x)=v(x)$), where the analogy to 3d scattering
theory is most plain.  These examples are not meant as models of real atoms or molecules,
but simply illustrate the general principles.

\subsection{Potential scattering: $N_T=0$}
We divide the illustrations into two cases, to make as clear as possible
the distinct physical principles involved.  We first treat potential
scattering, i.e., a single-particle scattering from a potential, and
show how box calculations yield the exact result.
Thus $N_T=0$, where $N_T$ is the number of particles in the target.
In the next section, we treat the
more complex interacting case, analogous to electron scattering from
H, where $N_T=1$.

\subsubsection{Phase shifts in 1d}
For a symmetric 1d potential, parity plays a role analogous to the partial waves
in 3d.  Thus, there are only two channels, symmetric and antisymmetric.  Careful
analysis allows definition of two phase shifts\cite{E65}:
\begin{equation}
e^{2i\delta^{\pm}}=t\pm r.
\end{equation}
The + sign corresponds to an even  continuum 
solution and the - sign to an odd solution. The amplitudes $t$ and $r$ are defined for scattering boundary conditions, i.e., the asymptotic wave functions are,
\begin{eqnarray}
\phi_{k}(x)\to &e^{ikx}+re^{-ikx}\quad &x\to-\infty\nonumber\\
\to& te^{ikx}\quad &x\to\infty.
\end{eqnarray}
The amplitudes $r$ and $t$ are related to the reflection and transmission coefficients, $R$ and $T$,
\begin{equation}
R=|r|^{2},~~~~~~T=|t|^2.
\end{equation}
A useful result from this analysis is
\ben
t = e^{i(\delta^++\delta^-)} \cos\Delta,
\label{e:tdelta1}
\een
where $\Delta=\delta^+-\delta^-$, showing that
resonances occur when $\Delta/\pi$ is an integer.

An example of a short-ranged central potential is the delta well potential, $v(x)=-Z\delta(x)$.
The free space example can be easily calculated, and the amplitudes $t$ and $r$ are given by
\begin{equation}\label{e:tdelta}
t=\frac{1}{1+Z/ik},~~~~~~~r=t-1.
\end{equation}
From this it follows that
\begin{equation}
e^{2i\delta^{+}}=\frac{1-Z/ik}{1+Z/ik},
\end{equation}
and we find for the phase shift
\begin{equation}\label{e:freephase}
\tan \delta^{+} = \frac{Z}{k}.
\end{equation}
The odd phase shift $\delta^{-}=0$.  (This is special to the delta potential, but is not
true in general.)

\subsubsection{Scattering solution in a box}
Now consider putting the potential inside an infinite well, with walls
located at $x=\pm L/2$. Since the potential is placed in a box, the states are discrete and there is only a discrete
set of allowed values for the eigen-energies $E_{n}^{\pm}$, $n=1,2,\ldots$. By comparing wavefunctions beyond the range
of the potential, one finds
the phase shifts corresponding to the even and odd solutions are given by
\begin{equation}\label{e:evenphase}
\delta^{\pm}=\frac{\pi}{4}-\frac{k_n^{\pm}L}{2}\pm\frac{\pi}{4},
\end{equation}
with $k_{n}^{\pm}=\sqrt{2E_{n}^{\pm}}$.

The even solution to the Schr{\"o}dinger equation in a box with boundaries at $x=\pm L/2$ is
\begin{equation}\label{e:phibox}
\phi_n^+(x) = A \sin{k_n^{+}\left(\frac{L}{2}-|x|\right)},
\end{equation}
with normalization constant
\begin{equation}\label{e:abox}
A=\sqrt{\frac{2}{L}}\left(1-\frac{\sin{k_n^{+}L}}{k_n^{+}L}\right)^{-1/2}.
\end{equation}
The discrete allowed $k_n^{+}$ values satisfy
\begin{equation}
k_n^{+}=Z\tan{\frac{k_n^{+}L}{2}}.
\end{equation}
(For the derivation, see for example Ref.~\cite{BBB08}.)
Taking the tangent of Eqs.~(\ref{e:evenphase}), we obtain for the phase shift,
\begin{equation}\label{e:boxphase}
\tan\delta^{+}=\cot{\frac{k_n^{+}L}{2}}=\frac{Z}{k_n^{+}}.
\end{equation}
\begin{figure}[tbp]
\includegraphics[width=7cm]{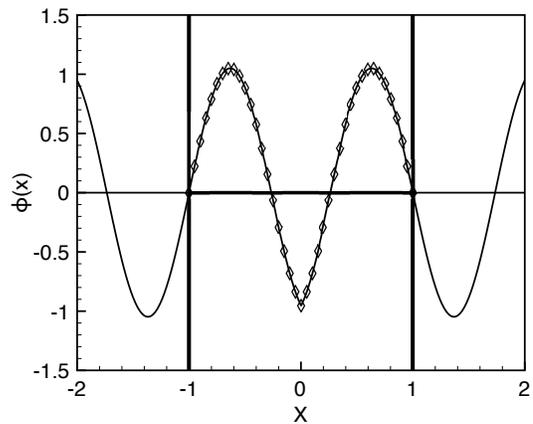}
\caption{\label{f:deltaorb} Solid line: even continuum state for a delta function with strength $Z=2$ at $k=4.2748$, corresponding to the first state of positive energy when walls with $L=2$ are imposed. Symbols: bound state for the same potential at the same energy in a box with walls at $x=\pm L/2 = \pm 1$.}
\end{figure}
Eqs.~(\ref{e:freephase}) and~(\ref{e:boxphase}) are identical,  but Eq.~(\ref{e:boxphase}) can only be evaluated at a finite number of energies $k_n^{+}$. In Fig.~\ref{f:deltaorb} we show both a continuum scattering state and a box state for the delta well potential at the same energy. From the figure it is clear that they are identical within the box, and therefore have the same phase shift.

The odd solution is
\begin{equation}\label{e:phioddbox}
\phi_n^{-}(x) = \sqrt{\frac{2}{L}}\sin{k_n^{-}x}.
\end{equation}
These are simply the particle-in-a-box wave functions; the delta potential has no effect on the odd states. The corresponding energies are the particle in a box energies,
\begin{equation}\label{e:eneroddbox}
k_n^{-} = \frac{2n\pi}{L}
\end{equation}
where $n=1,2,3,\ldots$. This corresponds to a phase shift $\delta^{-}=0$.

\subsubsection{DFT treatment}
It is now instructive to see how this one-electron example should be treated with (TD)DFT.
The first step in any linear-response TDDFT calculation is to find the
KS ground-state potential of the $(N_T+1)$-electron system.
 The ground-state KS potential is given by
\begin{equation}
v_{\rm KS}(x)=v_{\rm ext}(x)+v_{\rm Hxc}(x).
\end{equation}
For one electron ($N=1$), $v\x(\br)=-v\H(\br)$ and $v\c=0$,
and the KS potential is just the external potential. 
The KS orbitals for positive energies are equal to the scattering states
we obtained above in the exact case (with or without the box). 

Next one usually calculates
the TDDFT correction corresponding to the excitation of the electron
into one of the positive energy states. Again, for a one-electron system, $v\Hxc(\br,t)=0$, and thus $f_{\rm Hxc}(\br,\br',\omega)=0$.
From Eqs.~(\ref{e:omegas}) and~(\ref{e:kmat}) it follows that the
true excitation energies are equal to the KS orbital energy differences.
The TDDFT scattering phase shift is therefore equal to the exact phase shift in this trivial case. This also tells us the condition needed for any TDDFT treatment to recover the exact results for this case: Both the ground-state potential and the TDDFT kernel must be self-interaction free for one electron.

\subsection{Interacting case: $N_T=1$}
The case in the previous section was a trivial one for (TD)DFT, but checked
that our methodology worked in that limit.
Now we study a case that is not trivial.
We consider two fermions, one of which resides in the bound state of the 
delta potential well and the other scatters from this target.
The two electrons interact through a delta potential repulsion.
We show in this section that, with exact functionals,
the TDDFT results reproduce the exact results.

\subsubsection{Exact solution}
The Hamiltonian of this problem is given by~\cite{R71}
\begin{equation}
\hat H=-\frac{1}{2}\left(\frac{d^2}{dx_1^2}+\frac{d^2}{dx_2^2}\right)
-Z\left[\delta(x_1)+\delta(x_2)\right]+\lambda\delta(x_1-x_2)
\end{equation}
The two electrons interact via a delta-function repulsion potential, 
scaled by $\lambda$. In Ref.~\cite{WBb06} the scattering amplitudes
for both the exact and (TD)DFT case were derived to first-order in
$\lambda$. They found the transmission amplitude for the exact case~\cite{WBb06}.
For triplet scattering, the bound electron has no effect, and
\begin{equation}\label{e:ttripexact}
t_{\rm trip} = t^0 = \frac{ik}{Z+ik}
\end{equation}
identical to Eq.~(\ref{e:tdelta}).  But the singlet has a correction due to the
interaction:  
\begin{equation}\label{e:tsingexact}
t_{\rm sing} = t^0\, \left( 1  - \frac{2\, i\, \lambda\, k}{Z^2+k^2}\right).
\end{equation}
Our methods should be able to reproduce this leading correction.

\subsubsection{Exact ground-state KS potential}
 \begin{figure}[tbp]
  \begin{center}
   \includegraphics[width=7cm]{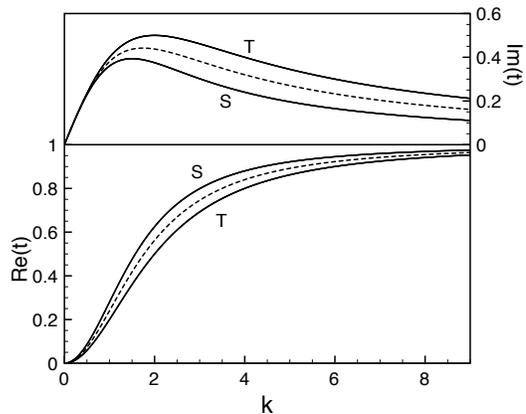}
   \caption{\label{f:transmiss_nobox}Real and imaginary parts of the exact (solid lines) and KS (dashed lines) transmission amplitude  for the delta potential well with contact interacting electrons. In this plot 
$Z=2$ and $\lambda=0.5$.}
  \end{center}
\end{figure}
To reproduce this result with TDDFT, first the ground-state KS problem must be solved.
 The ground-state KS potential of the $(N_T+1)$-electron system is given by
\begin{equation}
v_s(x)=-Z\delta(x)+\lambda|\phi_s(x)|^2
\end{equation}
to first-order in $\lambda$.  It is possible to solve the corresponding 
self-consistent Schr\"odinger equation exactly, finding
\begin{equation}
\phi_s(x)=\sqrt{Z}e^{-Z|x|}+\frac{\lambda}{8\sqrt{Z}}\left[
Ze^{-3Z|x|}+e^{-Z|x|}(4Z|x|-3)
\right]
\end{equation}
To first order in $\lambda$, the KS potential is therefore
\begin{equation}\label{e:scattKSpot}
v_s(x)=-Z\delta(x)+\lambda Z e^{-2Z|x|}.
\end{equation}
Transmission through this potential can easily be calculated using the
distorted-wave Born approximation, which is
exact to leading order in $\lambda$, producing the KS transmission amplitude~\cite{WBb06},
\begin{equation}\label{e:dwbats}
t_s = t^0\, \left( 1  - \frac{i\, \lambda\, k}{Z^2+k^2}\right).
\end{equation}
The transmission through the KS potential is exactly midway between 
singlet and triplet scattering.
In Fig.~\ref{f:transmiss_nobox}, we plot both the exact and KS transmission functions for $Z=2$ and $\lambda=1/2$ to first order in $\lambda$.

\subsubsection{TDDFT corrections}
The next step in a TDDFT calculation is to correct the KS transmission for both
singlet and triplet scattering, using the Hartree-exchange (Hx) kernel
as correlation effects are higher order in $\lambda$. Because we are treating both singlet and triplet scattering, we must use spin-TDDFT. This has no effect on the closed shell ground-state, but
the exact spin-dependent kernel is, to first order in $\lambda$,~\cite{PGG96}
\begin{equation}
f\Hx^{\sigma\sigma'}= \lambda\, \delta(x-x')(1-\delta_{\sigma\sigma'})
\end{equation}
where $\sigma$ indicates the spin.

Using this kernel the following amplitudes are obtained, by
applying a distorted-wave Born approximation directly to the Dyson-like
equation relating the true and KS response functions~\cite{WBb06}:
\begin{eqnarray}\label{e:tddftconts}
t_{\rm sing} &=& t_s - \lambda\frac{ik^2}{(k-iZ)^2(k+iZ)}\nonumber\\
&=&t^{0}(1-\frac{2i\lambda k}{Z^{2}+k^{2}})
\end{eqnarray}
\begin{eqnarray}
t_{\rm trip} = t_s + \lambda\frac{ik^2}{(k-iZ)^2(k+iZ)}=t^{0}
\end{eqnarray}
These equations are identical to the exact equations~\ref{e:tsingexact} and~\ref{e:ttripexact}.
This illustrates that TDDFT reproduces the exact answer.

\subsection{$N_T=1$ in a box}
We are finally ready to perform the most important test of our
methodology.  We use TDDFT with exact ground-state
and time-dependent functionals and with box boundary conditions
to reproduce the exact interacting scattering results above.

\subsubsection{KS ground-state in box}
To begin, we need first solve the ground-state KS problem in the box.
Now, the KS potential, Eq.~(\ref{e:scattKSpot}), is not zero beyond some radius $R$,
but extends out to $\infty$.  
But since it is exponentially decaying, we can choose
$L>>1/Z$, and have only exponentially small corrections to the
free results. Then the KS potential inside the box is simply given by Eq.~(\ref{e:scattKSpot}).
For the even states, we can use the eigenstates Eq. \ref{e:phibox} to
zero-order in $\lambda$, and then apply first-order perturbation theory.
The zeroth-order eigenstates simplify to,
\begin{equation}\label{e:evenKSboxstate}
\phi_n^+(x)=\sqrt{\frac{2}{L}} \sin{k_n^+\left(\frac{L}{2}-|x|\right)},
\end{equation}
i.e., the normalization constant of Eq. \ref{e:abox} simplifies, by using 
the allowed values of $k_n^+$ and the condition $L >> 1/Z$.  Note there is
no assumption concerning the relation between $k_n^+$ and $L$.

We next treat the interaction potential as a perturbation.
The energies of the scattering states in the box can be expressed as,
\begin{equation}
\varepsilon = \varepsilon_n + \delta\varepsilon_n
\end{equation}
where $\epsilon_n$ are the energies of the delta potential well without interaction. The perturbed energy is given by
\begin{equation}\label{e:KSpertener}
\delta\varepsilon_n=\left<n\right|\delta v\left|n\right>
= 2\lambda Z \int_{0}^{L/2} dx e^{-2Z|x|}| \phi_n(x) |^2 .
\end{equation}
An elementary integration yields:
\ben
\delta \varepsilon_n 
= \frac{\lambda}{L}\cdot\frac{1}{1+Z^2/k_n^{2}}
\een
for either odd or even states, using Eq.~(\ref{e:abox}) and 
$L >> 1/Z$.  Now, since $\delta k = \delta\varepsilon/k$,
\begin{equation}
L\delta k_{n} = L\frac{\delta v_n}{k_n} = \frac{\lambda}{k_n}\cdot\frac{1}{1+Z^2/k_n^2}.
\end{equation}
Finally, inserting these into Eqs.~(\ref{e:evenphase}) for the corrections to the phase shifts,
and then calculating the corresponding transmission amplitude from Eq.~(\ref{e:tdelta1}),
we obtain for the KS transmission
amplitude for two contact interacting electrons in a delta well potential,
\begin{equation}
t_s= t^0\, e^{-i\lambda/k_n(1+Z^2/k_n^2)}=t^{0}\left(1-\frac{i\lambda k_{n}}{Z^{2}+k_{n}^{2}}\right)
\end{equation}
agreeing with the exact result, Eq.~(\ref{e:dwbats}).

\subsubsection{TDDFT corrections in box}
 \begin{figure}[tbp]
  \begin{center} 
   \includegraphics[width=7cm]{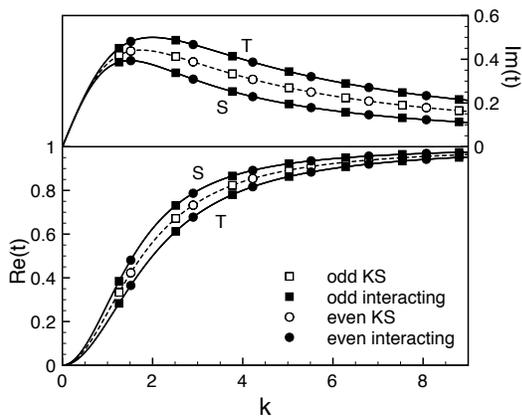}
   \caption{  \label{f:transmiss} Real and imaginary parts of the KS transmission amplitude and 
of the interacting singlet and triplet amplitudes for the delta potential well with contact interacting electrons. In this plot 
$Z=2$ and $\lambda=0.5$. For the results from the box (symbols) we used $L=5$.}
  \end{center}
\end{figure}
Last, we need to show that the TDDFT corrections for the box yield the
exact results.  We start from the $\Omega$-matrix of Eqs.~(\ref{e:omegas}) and~(\ref{e:kmat}). We first make the small matrix approximation (SMA) that assumes that only diagonal elements of this matrix contribute to the excitation energies.
For the singlet excitations, we can then write the $\Omega$-matrix in short-hand notation,
\ben
\Omega^+_q = \omega_q + \langle q | f\Hx | q \rangle,
\een
where $q=(0,n)$ indicating that we only consider transitions from the ground-state. Writing $\varepsilon_n=\Omega_q-I$, where $I$ is the ionization potential, we find:
\ben
\delta \varepsilon_{n} = 2\langle q | f\Hx | q \rangle,
\een
where
\bea
\left<q\right|f\Hx\left|q\right>
 &=& 2{\lambda Z}\int_{0}^{L/2}e^{-2Z|x|}|\phi_{n}(x)|^{2}dx,
\eea
i.e., the same as $\delta\varepsilon_{n}$ from Eq.~(\ref{e:KSpertener}). We then immediately obtain for the transmission coefficient,
\ben
t_{\rm sing}= t^{0}\left(1-\frac{2i\lambda k_{n}}{Z^{2}+k_{n}^{2}}\right).
\een
The same as in the continuum case Eq.~(\ref{e:tddftconts}). In the triplet case $f\Hx=0$ and the equations we obtain are identical to the continuum case, i.e. $t_{\rm trip}=t^{0}$. 

Thus exact TDDFT in the
box precisely reproduces both the TDDFT results without the box, and the exact
results from wavefunction theory, as shown in Fig.~\ref{f:transmiss}.

\section{Computational details}
 \label{comp}
Here we describe the details of the numerical calculations for real atoms and ions.
 We obtain the positive KS orbital energies (necessary to evaluate the phase shift in Eq.~(\ref{e:lrphase}))  from a well-established fully numerical
spherical DFT code~\cite{JE05}. This code includes many approximate xc-potentials, including the standard LDA and GGAs, and also the optimized effective potential method (OEP) for exact exchange calculations. The program does not use a basis set like most quantum chemistry codes, but instead works with a radial logarithmic grid. Both the KS orbital energies and the potentials are optimized in a self-consistent way.
The code is supplemented by the option to insert a hard-wall at a distance
$R_b$ from the origin. We always use a large number of grid points, $>1500$, to ensure convergence with the number of grid points.

The TDDFT excitation energies are calculated with our own code that solves the 
radial TDDFT equations. Since we are dealing with small systems,  we can exactly diagonalize the ${\Omega}$-matrix of Eq.~(\ref{e:omegas}) without regard to computational cost. The TDDFT code reads in the grid points, ground-state orbitals, and ground-state orbital energies from the DFT calculation. The number of states included in the ${\Omega}$-matrix are always chosen large enough for the transition energies to have reached convergence. 

Since we want to asses the performance of TDDFT kernels, we use a highly accurate ground-state xc-potential for the helium atom, obtained from very accurate wave functions, by Umrigar {\em et al.}~\cite{UG94}. We regard this as the essentially exact xc-potential. This potential is spline-fitted onto the grid points and imported into the code.

 \section{Results}
 \label{results}
 In this section we show our results for e-He${}^{+}$ and e-H scattering. In case of He we also study the quantum defect. We study in detail the effect of different xc-potentials and xc-kernels. We also study the effect of the location of the potential wall.
 
\subsection{Usefulness of the KS potential}
 \begin{figure}[tbp]
  \begin{center}
   \includegraphics[width=7cm]{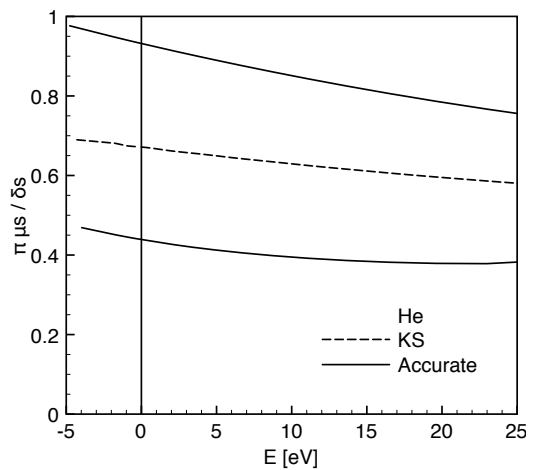}
   \caption{  \label{f:HeKSref} s-Wave phase shift for the He atom. Accurate data from Refs.~\cite{B02,D96}. All data below the KS result corresponds to singlet values, all data above to triplet values. For $E<0$, we plot $\pi$ times the QD, so it merges smoothly with the phase shift.}
  \end{center}
\end{figure}
In Fig.~\ref{f:HeKSref} we show highly accurate s-QD values (multiplied by $\pi$ in this and all following figures), $\mu_{s}$, and s-wave phase shifts, $\delta_{s}$ obtained from the literature~\cite{B02,D96}. We note, for our DFT readers, how smooth these curves are, demonstrating the continuity of eigenstates through $E=0$. All eigenenergies of the infinite Rydberg series can therefore be captured by a single smooth curve, as was already indicated in Eq.~(\ref{eq:fit}). The fit coefficients corresponding to the continuous curves in this and all following figures are available as supplementary material. 
Comparing our KS results to the literature data, we see that the KS QD and phase shifts are in-between the accurate wave function results. The KS results also mimic the correct shape of the accurate curves. We note for our non-DFT readers how much information is contained in the exact ground-state KS potential of the $(N_{T}+1)$-electron system (in this case He). All we need TDDFT to do is to shift the curve down in case of the singlet and upward in case of the triplet. Thus, an important message of this review for non-DFT readers is just how good a starting point the accurate KS potential is.

\subsection{TDDFT corrections}
To see how, and if, TDDFT corrects the KS results, we study the effect of the $f_{\rm xc}$ kernel that appears in Eq.~(\ref{e:kmat})
by using a sequence of approximations that include more and more physics. While there is as yet no explicit reverse-calculation for the exact kernel, we know that, if we had that exact kernel, it would reproduce the reference curves of Fig.~\ref{f:HeKSref}.

 \begin{figure}[tbp]
  \begin{center}
   \includegraphics[width=7cm]{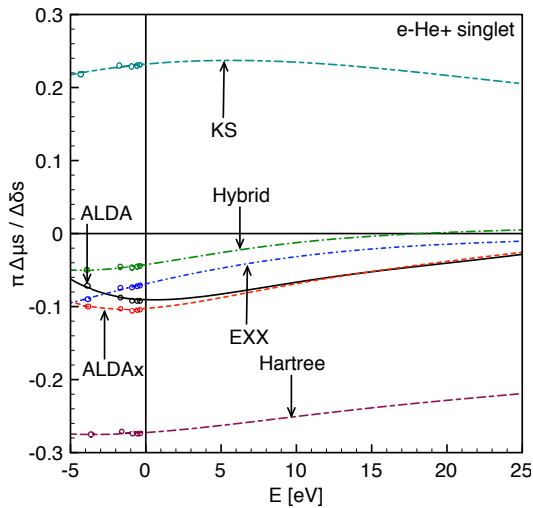}
   \caption{  \label{f:ErrorsSkernels} Error of calculated singlet s-wave phase shifts with respect to the reference data from Refs.~\cite{B02,D96}. For $E<0$, we plot $\pi$ times the QD error, so it merges smoothly with the phase shift error. Color figure online.}
  \end{center}
\end{figure}
 \begin{figure}[tbp]
  \begin{center}
   \includegraphics[width=7cm]{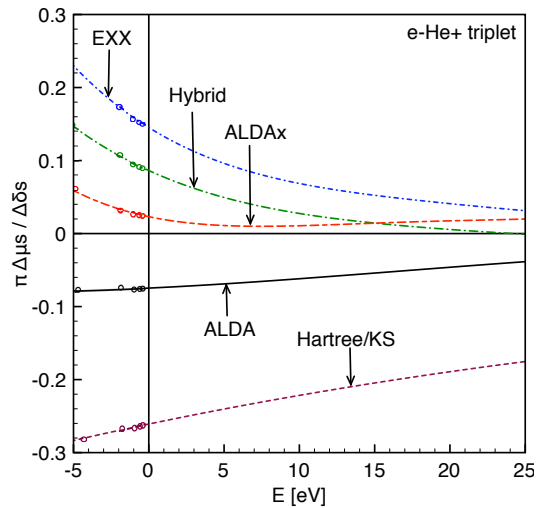}
   \caption{  \label{f:ErrorsTkernels} Same as Fig.~\ref{f:ErrorsSkernels}, but for triplet. Color figure online.}
  \end{center}
\end{figure}
In Figs.~\ref{f:ErrorsSkernels} and~\ref{f:ErrorsTkernels} we show the difference of the TDDFT results with the exact results for the s-wave quantum defect and phase shift of singlet and triplet respectively. The difference is calculated by fitting the results to a polynomial of order 2 or higher, and subtracting the exact values from the TDDFT values for the quantum defect and phase shift. In case of the quantum defect we indicate with circles the locations of the finite Rydberg states. We note again that for fitting just the QD, i.e. obtaining the $\mu_{i}$ coefficients, a polynomial of order 2 is enough, while we generally need a higher order polynomial to fit the QD and phase shifts together. We now analyze our results in much detail.

The first step in our analysis is to set the xc-kernel to zero. In case of the triplet, Eq.~(\ref{e:kmat}) tells us that we are left with just the KS orbital energy differences, i.e. there is no correction. For the singlet excitations there is a correction. The approximation of setting the xc-kernel to zero is called the Hartree approximation and is also known as the random phase approximation (RPA) in physics and does not involve any TDDFT. Fig.~\ref{f:ErrorsSkernels} shows that the Coulomb kernel corrects the KS result in the right direction, but overcorrects by about as much as the original error. We observe that even without an xc-kernel this methodology splits the bare KS results into separate singlet and triplet curves.

The next step is to include exchange, which usually dominates over correlation.
Since we are considering a two-electron system, the exact exchange kernel (EXX) is also known; it is given by Eq.~(\ref{eq:fexx}).
We see in Figs~\ref{f:ErrorsSkernels} and~\ref{f:ErrorsTkernels} that the Hartree results are considerably corrected by inclusion of the exchange kernels. In the triplet case the EXX correction is in the right direction, but it overcorrects.

Correlation is included by considering the full ALDA kernel. The ALDA kernel performs slightly worse than the EXX kernel in case of the singlet results. In case of the triplet results the ALDA performs better than EXX, especially at lower energies. The size of the error in the singlet and triplet full ALDA curves compared to the reference data is very similar. 

For completeness we also show ALDAx results, which includes only the exchange part of the ALDA kernel. In the singlet case, the full ALDA results merge with ALDAx for higher energies, the full ALDA improving on ALDAx for lower energies. In the triplet case the ALDAx results are better than full ALDA for all energies. 

The good results of the EXX kernel and the correlation correction of ALDA, suggests to combine the two. This leads to the hybrid kernel~\cite{BPG00} of Eq.~(\ref{eq:fhyb}). In the singlet case the hybrid kernel clearly improves the EXX and ALDA kernels over the entire energy range. One of the conclusions of Ref.~\cite{BPG00} was that the hybrid kernel improves the (bound-bound) excitation energies of He in almost all cases. On the continuum side, we see that the hybrid kernel performs the best of all these kernels in case of the singlet, and for the triplet it improves on EXX, but only improves the ALDA for large energies. Another conclusion of Ref.~\cite{BPG00} is that the hybrid kernel improves the bound-bound triplet $s$ excitation energies of He. Our finding that the ALDA performs better instead is most probably because the TDDFT calculation of Ref.~\cite{BPG00} only includes the lowest 34 unoccupied states of $s$ and $p$ symmetry in the calculation of the $\Omega$-matrix, whereas we included over 350 states including the continuum in all cases, notably improving the numerical accuracy of the TDDFT results. 

\subsection{Approximate ground-state potentials}
 Until now we have always used the exact ground-state potential for our calculations. It is interesting to study the influence of the ground-state potential on the TDDFT results as, in practice, only approximate potentials are available. There is a wealth of such potentials and we cannot cover all of them in this paper. 
 
For illustrative purposes we therefore limit ourselves to two potentials: the most simple potential of all, the local density approximation (LDA), and the orbital-dependent exact exchange potential implemented within the optimized effective potential scheme (OEP). We expect little qualitative change moving from LDA to GGA, or adding GGA correlation to EXX. We showed these potentials together with the exact potential in Fig.~\ref{f:He_LDAexact}. From the figure we see that the LDA is very different from the exact potential. The potential is too narrow in the interior region, resulting in a bound state of too high energy, and it is short-ranged, completely missing the $-1/r$ tail. But the shape of the LDA in the interior region is typically very good. It is almost equal to the exact potential, shifted upward by a constant, therefore occupied LDA orbitals yield a remarkably accurate density, except in the tail region. This feature can be used to extract a highly accurate QD from the LDA potential~\cite{WB05}.

On the other hand, the EXX potential is almost indistinguishable from the exact potential; it is very similar in the interior region and has the correct asymptotic behavior. The exact potential is slightly deeper than the OEP potential near the nucleus (see the inset in Fig.~\ref{f:He_LDAexact}). We expect the results from the OEP potential to be very close to the exact results.

 \begin{figure}[tbp]
  \begin{center}
   \includegraphics[width=7cm]{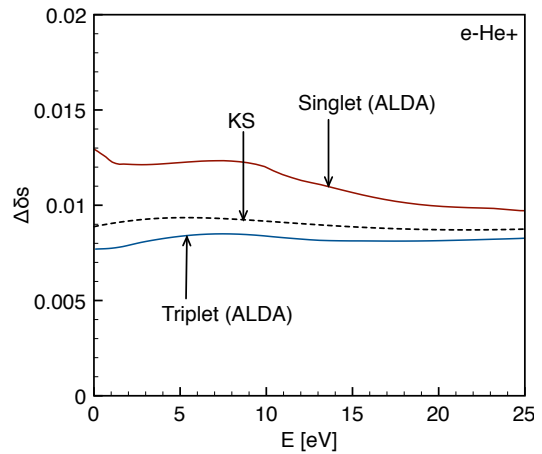}
   \caption{  \label{f:ErrorOEPexact} The error in the s-wave phase shifts obtained from the EXX potential, relative to those from the exact KS potential. Color figure online.}
  \end{center}
\end{figure} 
We show the error in the phase shift obtained using the EXX ground-state potential in Fig.~\ref{f:ErrorOEPexact} (using the ALDA kernel). The LDA results diverge completely and there is no quantum defect because of the missing Rydberg series, therefore we do not show these results in the figure. The EXX  results are shifted a little bit upwards compared to the exact results, but this error is an order of magnitude smaller than those due to approximate kernels.

We conclude that using EXX potentials is clearly sufficient for these calculations. On the other hand, the LDA results diverge. Thus, without the special handling of Ref.~\cite{WB05}, standard approximations (LDA, GGAs, hybrids) are useless here.

 \subsection{Effect of the wall}
 \begin{figure}[tbp]
  \begin{center}
   \includegraphics[width=7cm]{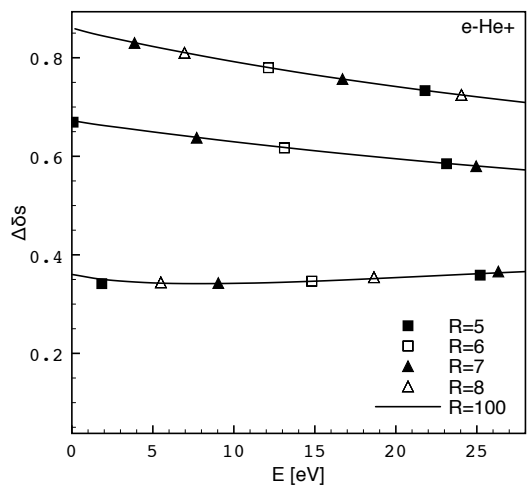}
   \caption{  \label{f:Hewalls} Exact KS (middle) and TDDFT-corrected (ALDA) results as a function of the wall radius. The singlet is above, and the triplet below, the KS curve..}
  \end{center}
\end{figure}
In this section, we demonstrate our claim that the infinite wall can be placed quite close to the atom, as long as the part of the potential that differs from $-1/r$ is essentially zero beyond the wall (Eq.~(\ref{eq:coulsr})). In Fig.~\ref{f:Hewalls}, we show the singlet and triplet $s$-wave phase shifts for wall distances between 5 and 8 bohr, and compare the results with $R_b=100$ a.u.
The smallest distance is determined by the requirement that the box does not significantly distort the self-consistent potential in the region of significant density, i.e., occupied orbitals must be unaffected.
The results for the small wall distances indeed correspond to the correct phase shifts. By varying the wall distance at small $R_{b}$ we can fill out the complete curve, giving an alternative to performing a single calculation with a very large $R_b$. This is an important conclusion if one wants to use finite basis-set methods instead of a grid based code~\cite{F08}, because the range of finite basis-sets, even if they include very diffuse functions, is limited to much shorter radii.

\subsection{Different $l$-values}
 \begin{figure}[tbp]
  \begin{center}
   \includegraphics[width=7cm]{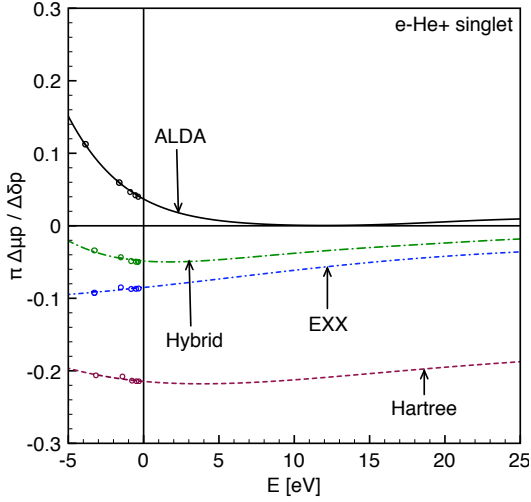}
   \caption{  \label{f:ErrorpSkernels} Same as Fig.~\ref{f:ErrorsSkernels}, for $p$-wave. Color figure online.}
  \end{center}
\end{figure}
 \begin{figure}[tbp]
  \begin{center}
   \includegraphics[width=7cm]{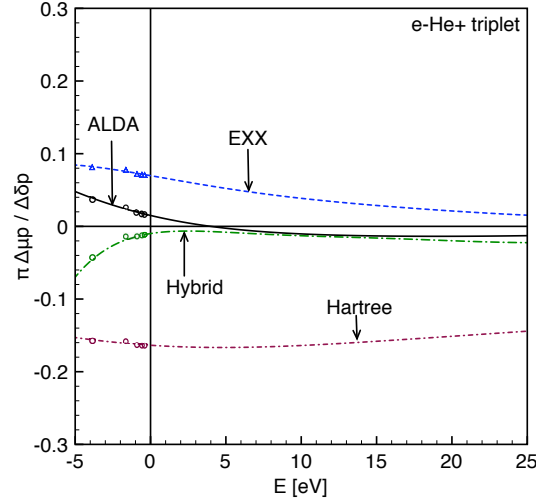}
   \caption{  \label{f:ErrorpTkernels} Same as Fig.~\ref{f:ErrorsTkernels}, for $p$-wave. Color figure online}
  \end{center}
\end{figure}
Until now we focussed on $s$-wave scattering and $s$-QDs. Of course we can also study higher $l$-values, using the $l$-dependent Eq.~(\ref{e:lrphase}) for the phase shift. We show the difference between our $p$-QD and phase shift (i.e. $l=1$) and the reference values  for He in Figs.~\ref{f:ErrorpSkernels} and~\ref{f:ErrorpTkernels}. In Ref.~\cite{FKB08}, we also showed results for $d$-wave scattering.

Adding xc-kernels notably improves the Hartree results just as in case of the $s$-QD and $s$-wave phase shift. For negative and low energies, the hybrid performs the best of the three kernels, but for larger energies ALDA performs better, being very close to the reference results. The bare EXX kernel does not perform as well in this case, unlike for $s$-wave phase shifts. Thus ALDA is better overall, and can easily be applied to any system tractable with ground-state DFT. 

\subsection{Scattering from a neutral system: $e$-H scattering}
 \begin{figure}[tbp]
  \begin{center}
   \includegraphics[width=7cm]{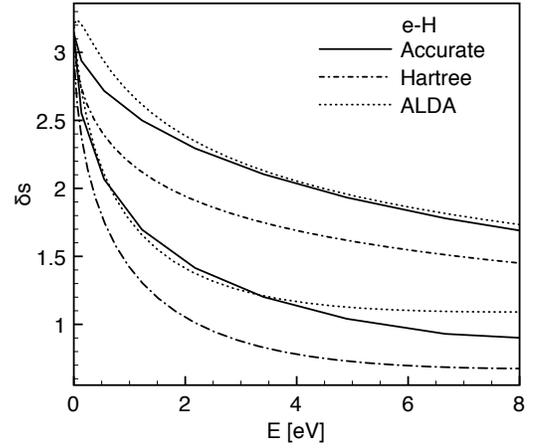}
   \caption{\label{f:eH} s-QD and s-wave phase shift for Hartree and ALDA compared to accurate data~\cite{S61}. The bottom three curves correspond to singlet, the top three to triplet.}
  \end{center}
\end{figure}
When we scatter from a positive ion like He${}^{+}$ as we did throughout this paper, the $(N_{T}+1)$-electron system is a neutral atom with a long-ranged KS potential. In Ref.~\cite{FWEZ07} we studied another two-electron system, namely electron scattering from the hydrogen atom. In this case the $(N_{T}+1)$-electron system, H${}^-$, is a negative ion with a short-ranged KS potential. Because of this, there is no Rydberg series or QD and only one bound state. We now calculate the phase shift using Eq.~(\ref{e:srphase}), i.e. relative to free potentials.

 We show our results in Fig.~\ref{f:eH}. We see that, just as for $e$-He${}^+$ scattering, the KS result, which is equal to triplet Hartree, lies in between both the accurate singlet and triplet values. ALDA corrects this and the ALDA results are again very close to the reference data. So TDDFT works well for electron scattering from both cations and neutral atoms. For more details on this particular system we refer to Ref.~\cite{FWEZ07}.

\section{Conclusions}
Electron scattering from H (as shown in Fig.~\ref{f:eH}) and electron scattering from He${}^{+}$ (as outlined in this paper) are two-electron systems. But our TDDFT methods can also be applied to elastic electron-scattering for systems with more than two electrons. 

Results for electron scattering from Li can be found in Ref.~\cite{FKB08}. The $(N_{T}+1)$-electron system is Li${}^{-}$ and has a short-ranged potential. The singlet TDDFT p-wave phase shift in this system suffers from an artifact that appears very close to zero energy (energies smaller than 1 eV). But for higher energies the TDDFT values correspond very well to the available reference data, as do the triplet p-wave and s-wave phase shifts. 

A many-electron system with a long-ranged potential is Be, corresponding to electron scattering from Be${}^{+}$. Scattering phase shifts of $e$-Be${}^{+}$  can be found in Ref.~\cite{FBb08}. In this system there are a number of low-energy resonances of double excitation character, not captured by our TDDFT method. Such effects require going beyond the adiabatic approximation of TDDFT and using a frequency dependent kernel~\cite{MZCB04}. But apart from missing these resonances TDDFT again performs well~\cite{FBb08}.

So in general this method is expected to work well for low-energy electron-atom scattering problems, without resonances due to multiple excitations, where the $(N_T+1)$-electron system is bound. We have not yet addressed the complications of non-spherical systems or of open-shells.

Another point we have not addressed are the oscillator strengths that correspond to the TDDFT excitation energies and how they behave when moving across the continuum. This situation is studied in Ref.~\cite{YB08}.  It is found that the KS oscillator strength corresponding to the ionization energy (exactly given by minus the energy of the KS HOMO) does not correspond to the exact oscillator strength. But the exact and KS oscillator strengths do merge at higher energies and several results for 2-electron systems are given in the reference.

In this paper we reviewed our theory and calculations for obtaining both the quantum defect and electron-atom scattering phase shifts using TDDFT. We have shown that our method of putting the system in a finite hard wall cavity in order to extract the continuum states and phase shifts works within TDDFT.
In our previous work we showed that the method works well for the short-ranged potential system $e$-H${}^{-}$. In this paper we first illustrated our box method on a (relatively) simple toy problem, in which the answer was known exactly. Next, we showed that the method works equally well for the long-ranged helium atom and $e$-He${}^{+}$ scattering. We also showed that the TDDFT quantum defect and phase shifts transition smoothly across the zero energy boundary, in accordance with Seaton's theorem.

Our final conclusion is that TDDFT works well for high-lying excitations in neutral atoms and cations. To infinity, and beyond!

MvF acknowledges The Netherlands Organization for Scientific Research (NWO) for support through a VENI grant.  KB thanks NSF for Grant No. CHE-0809859 and DOE DE-FG02-08ER46496. We thank Neepa Maitra for many useful discussions.
We also acknowledge the exact H$^-$ and He KS potentials from Cyrus Umrigar.

%

\begin{thebibliography}{10}

\bibitem{HK64}
P.~Hohenberg and W.~Kohn, {\em Phys. Rev.}, 1964, {\bf 136}(3), B864.

\bibitem{KS65}
W.~Kohn and L.~J. Sham, {\em Phys. Rev.}, 1965, {\bf 140}(4), A1133.

\bibitem{DG90}
R.~M. Dreizler and E.~K.~U. Gross, {\em Density Functional Theory: An Approach
  to the Quantum Many Body Problem}, Springer-Verlag, Berlin, 1990.

\bibitem{L82}
M.~Levy, {\em Phys. Rev. A}, 1982, {\bf 26}, 1200--1208.

\bibitem{RG84}
E.~Runge and E.~K. Gross, {\em Phys. Rev. Lett.}, 1984, {\bf 52}(12), 997.

\bibitem{EBF07}
P.~Elliott, F.~Furche, and K.~Burke in {\em Reviews of Computational
  Chemistry}, ed. K.~B. Lipkowitz and T.~R. Cundari;
\newblock Wiley, Hoboken, NJ, 2009.

\bibitem{G94}
T.~F. Gallagher, {\em Rydberg Atoms}, Cambridge University Press, Cambridge,
  1994.

\bibitem{CPW05}
R.~C\^{o}t\'{e}, T.~Pattard, and M.~Weidem\"{u}ller, {\em J. Phys. B: At. Mol.
  Opt. Phys.}, 2005, {\bf 38}(2).

\bibitem{F81}
U.~Fano, {\em Comments Atom Mol. Phys.}, 1981, {\bf 10}, 223.

\bibitem{D96}
G.~W.~F. Drake in {\em Atomic, Molecular, and Optical Physics Handbook}, ed.
  G.~W.~F. Drake;
\newblock AIP Press, Woodbury, NY, 1996;
\newblock p. 154.

\bibitem{NIST}
Y.~Ralchenko, F.-C. Jou, D.~Kelleher, A.~Kramida, A.~Musgrove, J.~Reader,
  W.~Wiese, and K.~Olsen, {\em NIST Atomic Spectra Database}, version 3.0.

\bibitem{FB06}
M.~van Faassen and K.~Burke, {\em J. Chem. Phys.}, 2006, {\bf 124}, 094102.

\bibitem{FBb06}
M.~van Faassen and K.~Burke, {\em Chem. Phys. Lett}, 2006, {\bf 431}, 410.

\bibitem{F06}
M.~van Faassen, {\em Int. J. Quant. Chem.}, 2006, {\bf 106}, 3235.

\bibitem{B02}
A.~Bhatia, {\em Phys. Rev. A}, 2002, {\bf 66}.

\bibitem{S58}
M.~J. Seaton, {\em Mon. Not. R. Astron. Soc.}, 1958, {\bf 188}, 504.

\bibitem{F98}
H.~Friedrich, {\em Theoretical Atomic Physics}, Springer-Verlag, 2 ed., 1998.

\bibitem{F35}
U.~Fano, {\em Nuovo Cimento}, 1935, {\bf 12}, 154.

\bibitem{FWEZ07}
M.~van Faassen, A.~Wasserman, E.~Engel, F.~Zhang, and K.~Burke, {\em Phys. Rev.
  Lett.}, 2007, {\bf 99}, 043005.

\bibitem{BH72}
U.~von Barth and L.~Hedin, {\em J. Phys. C}, 1972, {\bf 5}(13), 1629--1642.

\bibitem{PR72}
M.~M. Pant and A.~K. Rajagopal, {\em Solid State Commun.}, 1972, {\bf 10}(12),
  1157--1160.

\bibitem{UG94}
C.~J. Umrigar and X.~Gonze, {\em Phys. Rev. A}, 1994, {\bf 50}, 3827--3837.

\bibitem{PPLB82}
J.~P. Perdew, R.~G. Parr, M.~Levy, and J.~L. Balduz, {\em Phys. Rev. Lett.},
  1982, {\bf 49}(23), 1691--1694.

\bibitem{B88}
A.~D. Becke, {\em Phys. Rev. A}, 1988, {\bf 38}, 3098.

\bibitem{LYP88}
C.~Lee, W.~Yang, and R.~G. Parr, {\em Phys. Rev. B}, 1988, {\bf 37}(2),
  785--789.

\bibitem{PBE96}
J.~P. Perdew, K.~Burke, and M.~Ernzerhof, {\em Phys. Rev. Lett.}, 1996, {\bf
  77}(18), 3865--3868.

\bibitem{B93}
A.~D. Becke, {\em J. Chem. Phys.}, 1993, {\bf 98}, 5648--5652.

\bibitem{MSSP89}
B.~Miehlich, A.~Savin, H.~Stoll, and H.~Preuss, {\em Chemical Physics Letters},
  1989, {\bf 157}(3), 200 -- 206.

\bibitem{E03}
E.~Engel in {\em A Primer in Density Functional Theory}, ed. C.~Fiolhais,
  F.~Noguiera, and M.~A.~L. Marques;
\newblock Springer, New York, 2003;
\newblock p.~56.

\bibitem{TS76}
J.~D. Talman and W.~G. Shadwick, {\em Phys. Rev. A}, 1976, {\bf 14}, 36.

\bibitem{KLI92}
J.~B. Krieger, Y.~Li, and G.~J. Iafrate, {\em Phys. Rev. A}, 1992, {\bf 45}(1),
  101--126.

\bibitem{G96}
A.~G\"orling, {\em Phys. Rev. B}, 1996, {\bf 53}(11), 7024--7029.

\bibitem{C95}
M.~E. Casida in {\em Recent Advances in Density-Funtional Methods}, ed. D.~P.
  Chong;
\newblock World Scientific, Singapore, 1995;
\newblock p. 155.

\bibitem{PGG96}
M.~Petersilka, U.~J. Gossmann, and E.~K.~U. Gross, {\em Phys. Rev. Lett.},
  1996, {\bf 76}(8), 1212--1215.

\bibitem{VOC99}
I.~Vasiliev, S.~\"O\ifmmode~\breve{g}\else \u{g}\fi{}\"ut, and J.~R.
  Chelikowsky, {\em Phys. Rev. Lett.}, 1999, {\bf 82}(9), 1919--1922.

\bibitem{B06}
A.~Bhatia, {\em Phys. Rev. A}, 2006, {\bf 73}.

\bibitem{GSGB99}
S.~J.~A. van Gisbergen, P.~R.~T. Schipper, O.~V. Gritsenko, E.~J. Baerends,
  J.~G. Snijders, B.~Champagne, and B.~Kirtman, {\em Phys. Rev. Lett.}, 1999,
  {\bf 83}(4), 694--697.

\bibitem{MWY03}
P.~Mori-Sanchez, Q.~Wu, and W.~T. Yang, {\em J. Chem. Phys.}, 2003, {\bf 119},
  11001--11004.

\bibitem{KKP04}
S.~K{\"u}mmel, L.~Kronik, and J.~P. Perdew, {\em Phys. Rev. Lett.}, 2004, {\bf
  93}(21), 213002.

\bibitem{FBLB02}
M.~van Faassen, P.~L. de~Boeij, R.~van Leeuwen, J.~A. Berger, and J.~G.
  Snijders, {\em Phys. Rev. Lett.}, 2002, {\bf 88}(18), 186401--1--4.

\bibitem{BPG00}
K.~Burke, M.~Petersilka, and E.~K.~U. Gross in {\em Recent Advances in Density
  Functional Methods}, ed. V.~Barone, P.~Fantucci, and A.~Bencini, Vol. ~3;
\newblock World Scientific, 2000;
\newblock pp. 67--79.

\bibitem{E65}
J.~H. Eberly, {\em American Journal of Physics}, 1965, {\bf 33}(10), 771--773.

\bibitem{BBB08}
N.~Bera, K.~Bhattacharyya, and J.~K. Bhattacharjee, {\em American Journal of
  Physics}, 2008, {\bf 76}(3), 250--257.

\bibitem{R71}
C.~M. Rosenthal, {\em J. Chem. Phys.}, 1971, {\bf 55}(5), 2474--2483.

\bibitem{WBb06}
A.~Wasserman and K.~Burke, {\em Lect. Notes Phys.}, 2006, {\bf 706}, 493--505.

\bibitem{JE05}
H.~Jiang and E.~Engel, {\em J. Chem. Phys.}, 2005, {\bf 123}(22), 224102.

\bibitem{WB05}
A.~Wasserman and K.~Burke, {\em Phys. Rev. Lett.}, 2005, {\bf 95}, 163006.

\bibitem{F08}
M.~van Faassen, {\em in preparation.}

\bibitem{FKB08}
M.~van Faassen, M.~Koentopp, and K.~Burke, {\em to be submitted}.

\bibitem{S61}
C.~Schwarz, {\em Phys. Rev.}, 1961, {\bf 124}(5), 1468.

\bibitem{FBb08}
M.~van Faassen and K.~Burke, {\em in preparation}.

\bibitem{MZCB04}
N.~Maitra, F.~Zhang, R.~Cave, and K.~Burke, {\em J. Chem. Phys.}, 2004, {\bf
  120}, 5932.

\bibitem{YB08}
Z.~Yang and K.~Burke, {\em in preparation}.

\end{thebibliography}


\end{document}